# WAVELET ANALYSIS OF GALACTIC ROTATION CURVES

M. Kuassivi

BP 170374 Cotonou, Benin

## ABSTRACT

The spatial wavelet spectra of 73 published spiral galaxies's rotation curves are computed and their associated scaleograms are presented. Scaleograms are used to detect and isolate local features observed in spiral galaxies's rotation curves. Although wiggles and bumps are usually interpreted as signs of recent and on—going merging, the analysis of the scaleograms reveals regular patterns consistent with the presence of large—scale modes throughout the disk.

## 1. INTRODUCTION

In recent years, a great deal of focus has been put on the understanding of merger events in the evolution of galaxies. Our own galaxy has a tiny satellite galaxy (the Sagittarius Dwarf Ellitptical Galaxy) which is currently being ripped up and eaten by the Milky Way. It is thought these kind of events may be quite common in the evolution of large galaxies. The Sagittarius Dwarf galaxy is orbiting our galaxy at almost a right angle to the disk. It is currently passing through the disk; stars are being stripped off of it with each pass and joining the halo of our galaxy. There are other examples of these minor accretion events, and it is likely a continual process for many galaxies. Such mergers provide gas and dark matter to galaxies. Evidence for this process is often observable as warp and streams coming out of galaxies.

Most of the Early type galaxies show clear sign of interaction and tidal distortion with some of them having bumps and wiggles in the rotation curve (Struve & Conway, 2010). In such galaxies, the rotation curves curves are fitted best by models with scaled H I disks suggesting that a connection exists between large scale streaming motion and HI density distribution. As a matter of facts, discontinuities observed within the rotation curves desserve some attention. Analysis of these features is made easy by the use of the wavelet transforms.

The wavelet analyses properties of waves of various scale in real space (Jones 2009). Wavelet transforms are now being adopted for a vast number of applications, often replacing the conventional Fourier fransform. Many areas of physics have seen this paradigm shift, including molecular dynamics, astrophysics, density matrix localisation, seismology, optics, turbulence and quantum mechanics. Wavelet transforms have advantages over traditional Fourier transforms for representing functions that have discontinuities and sharp peaks, and for accurately deconstructing and reconstructing, non periodic and non stationnary sinals.

The goal of this paper is to use the wavelet transform for the analysis of perturbations in rotation curves.

In section 2. I describe the data and the method. In section 3. the results are briefly discussed. Scaleograms are presented in the annexe.

## 2. METHODOLOGY

The 97 rotation curves data used in this work come from the paper of Brownstein

& Moffat (2006). Using a short script written in IDL language, I manually extracted the data points from the numerical paper version. Table 1. shows the list of the galaxies finally used in this work. Then, I interpolated the rotation curves on a 0.1 kpc grid and substracted to each rotation curve the Newtonian model published by Brownstein & Moffat (using the same IDL script). Fig.1 is an example of the resulting output for F563—1 showing remarkable similarities with dissipative oscillations. Note that F563-1 is a low surface brightness (LSB) galaxy dominated by dark matter all the way into its center.

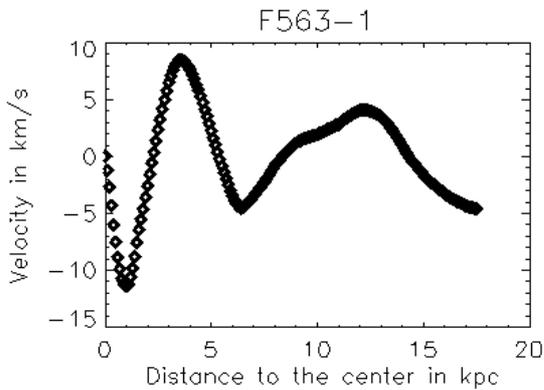

**Fig 1.** Reduced rotation curve of F563-1.

Wavelet transform is applied using the D.O.G mother because it ressembles the most bumps often observed. Note that the wavelet software was provided by C. Torrence and G. Compo, and is available at http://paos.colorado.edu/research/wavelets.

| Names | Vo (km/s) |
|---|---|
| DD0170 | 61 |
| F563-1 | 95 |
| F568-3 | 109 |
| F571-8 | 145 |
| F583-1 | 95 |
| NGC55 | 85 |
| NGC224 | 260 |
| NGC247 | 111 |
| NGC253 | 200 |
| NGC300 | 102 |
| NGC598 | 112 |
| NGC660 | 145 |
| NGC801 | 235 |
| NGC891 | 199 |
| NGC1097 | 290 |
| NGC1365 | 240 |
| NGC1417 | 150 |
| NGC1560 | 75 |
| NGC2403 | 135 |
| NGC2708 | 230 |
| NGC2841 | 310 |
| NGC2903 | 200 |
| NGC2998 | 190 |
| NGC3034 | 85 |
| NGC3079 | 200 |
| NGC3109 | 70 |
| NGC3198 | 155 |
| NGC3495 | 150 |
| NGC3521 | 200 |
| NGC3628 | 210 |
| NGC3672 | 220 |
| NGC3726 | 155 |
| NGC3877 | 161 |
| NGC3893 | 180 |
| NGC3972 | 129 |
| NGC4051 | 165 |
| NGC4088 | 170 |
| NGC4096 | 118 |
| NGC4100 | 179 |
| NGC4138 | 160 |
| NGC4157 | 189 |
| NGC4183 | 111 |
| NGC4217 | 190 |
| NGC4303 | 150 |
| NGC4321 | 260 |
| NGC4389 | 119 |
| NGC4527 | 180 |
| NGC4565 | 250 |
| NGC4736 | 150 |
| NGC4945 | 169 |
| NGC5033 | 210 |
| NGC5055 | 199 |
| NGC5457 | 211 |
| NGC5533 | 290 |
| NGC5585 | 85 |
| NGC5907 | 250 |
| NGC6503 | 119 |
| NGC6674 | 270 |
| NGC6946 | 160 |
| NGC6951 | 192 |
| NGC7331 | 250 |
| UGC2259 | 89 |
| UGC6399 | 86 |
| UGC6614 | 190 |
| UGC6667 | 85 |
| UGC6818 | 74 |
| UGC6917 | 102 |
| UGC6923 | 87 |
| UGC6930 | 110 |
| UGC6973 | 175 |
| UGC6983 | 110 |
| UGC7089 | 71 |

**Table.1** list of spiral galaxies used in this work with terminal velocities Vo.

3. CONCLUSION

Scaleograms allow to detect and characterize local perturbations in the rotation curves. We can see a variety of spatial frequencies among the saleograms (from 1 kpc$^{-1}$ to 40 kpc$^{-1}$). Such diversity is well expected if we are observing relief traces of merger events. Surprisingly enough, whithin each scaleogram the range of the detected frequencies is narrow and the amplitudes of the detection peaks are similar. Some scaleograms show a regular distributions of peaks that could be interpreted as trains of mergers.

However, such regular distributions are more likely related to harmonic modes in the rotation curves. I wish to point out that if a large—scale harmonic velocity structure is embedded in the rotation curve it may indeed be related to or be at the origin of the spiral arm structure. This would be consistent with the presence of modes in amorphous LSB galaxies such as F563-1.

As a conclusion, the interpretation of rotation curve discontinuities as arising from random mergers is put at a test.

ANNEXE

The figures below are the scaleograms obtain from Moffat & Brownstein (2006) rotation curves. Prior to the wavelet transformation, the Newtonian Model has been subtracted from the data. The X coordinates are the distance to the center of the galaxy in kpc. The Y coordinates are the spatial frequencies (in kpc$^{-1}$).

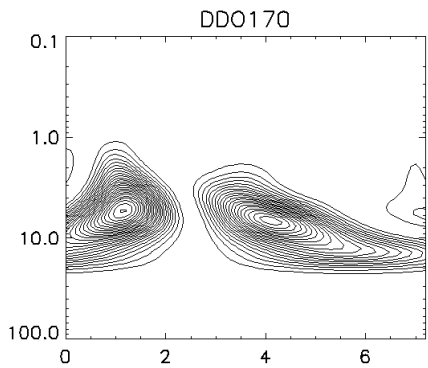
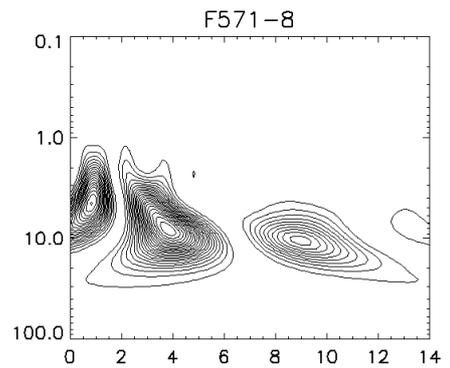
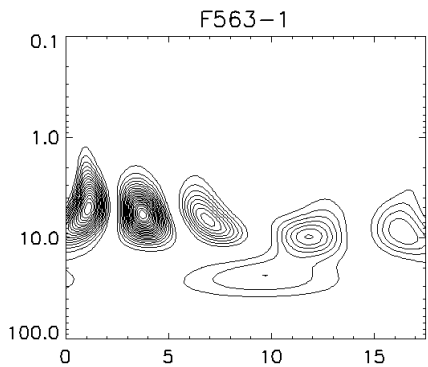
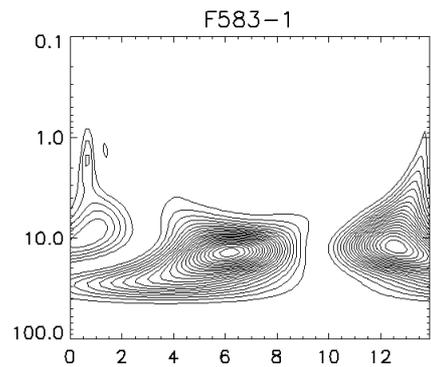
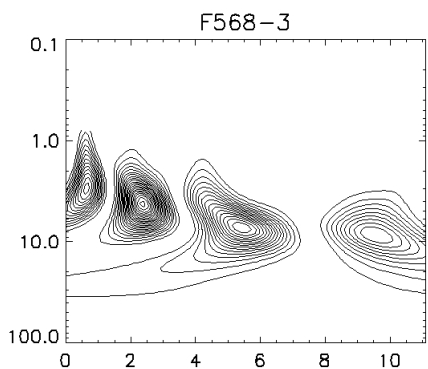
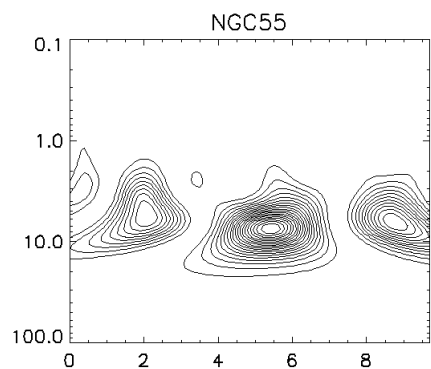

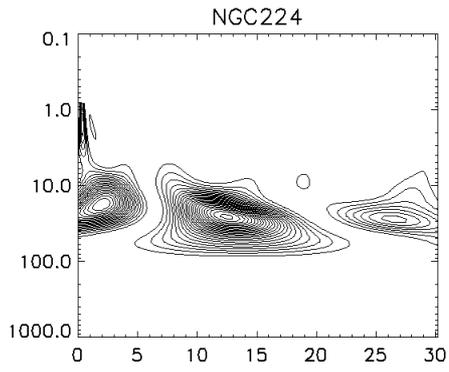
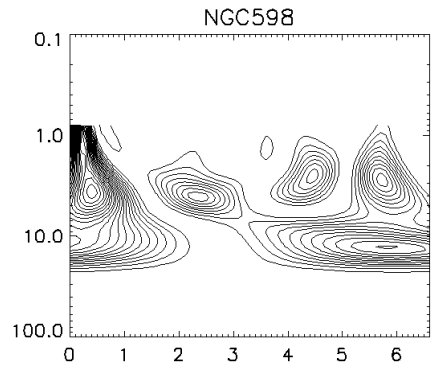
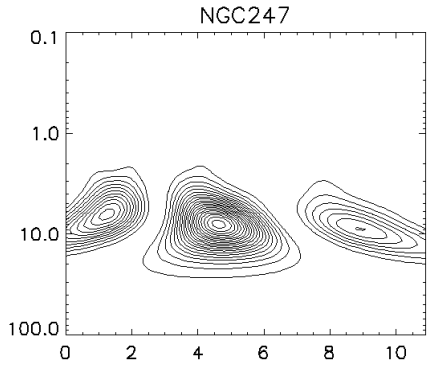
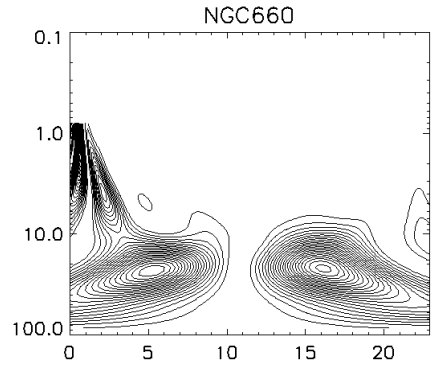
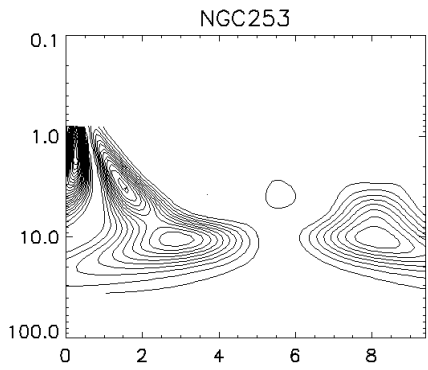
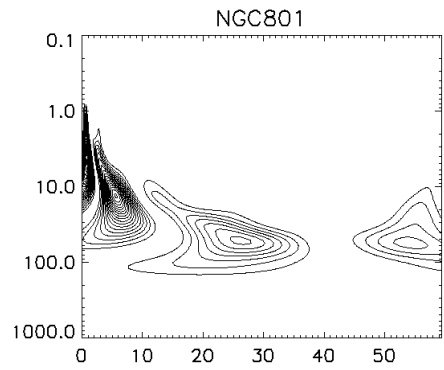
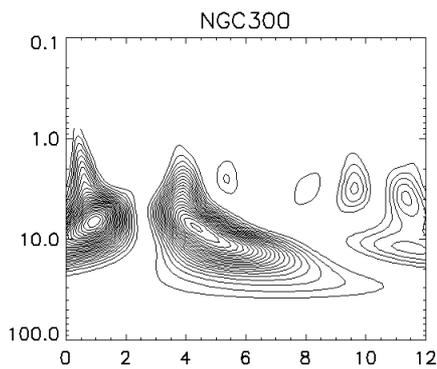
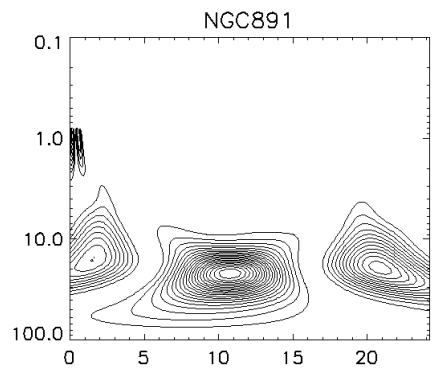

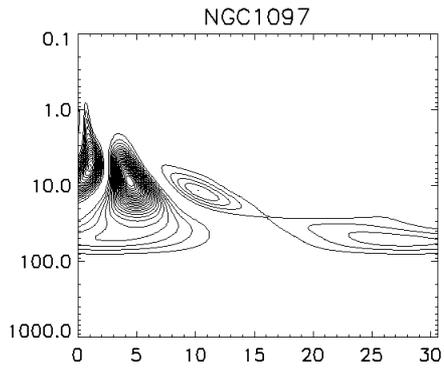
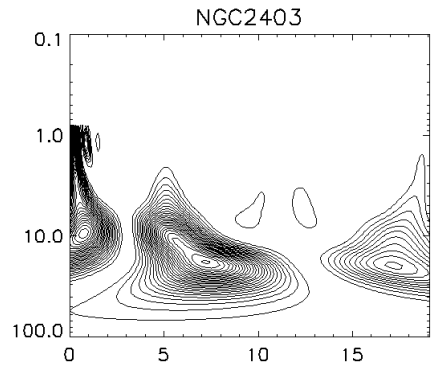
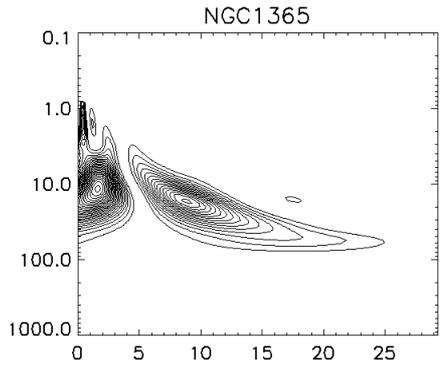
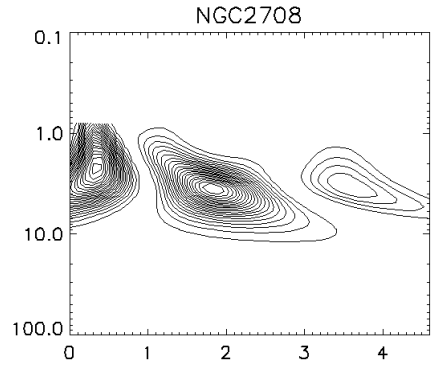
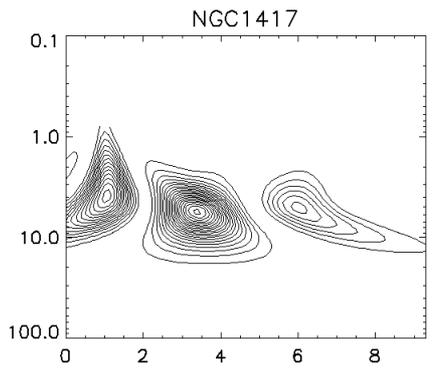
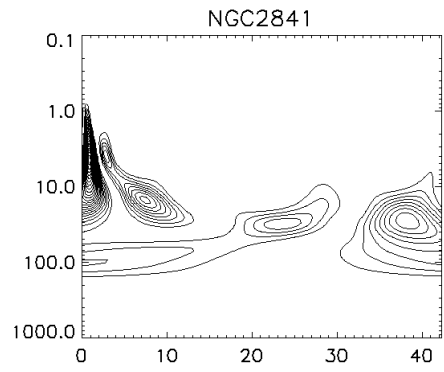
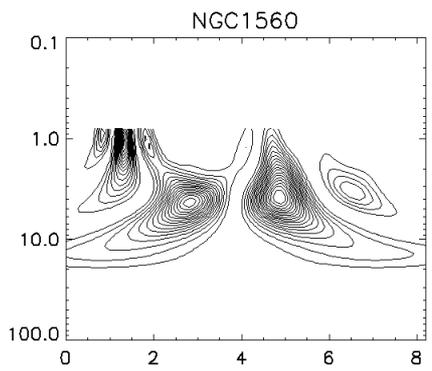
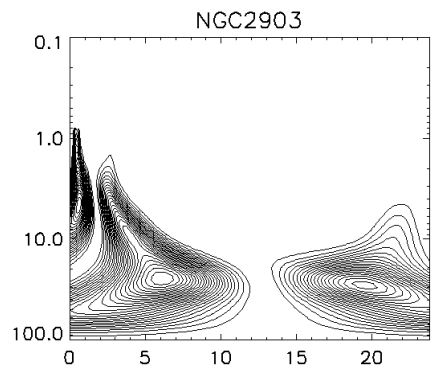

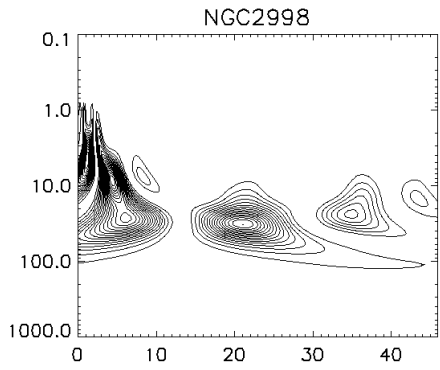
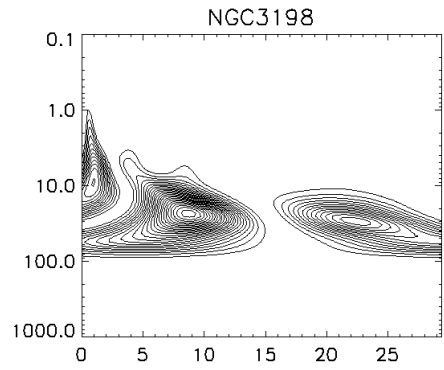
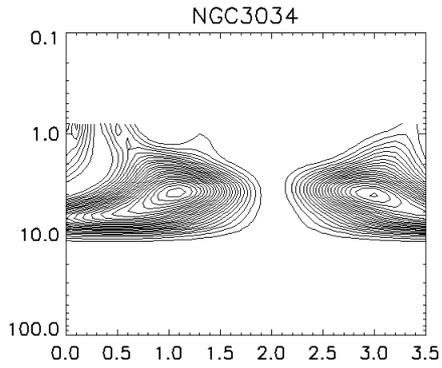
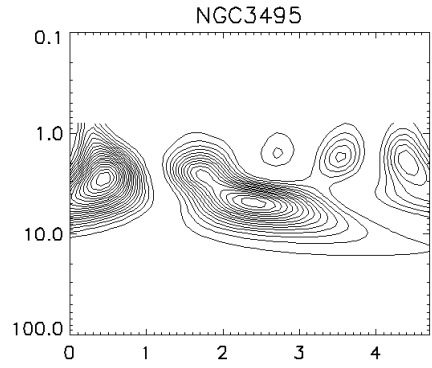
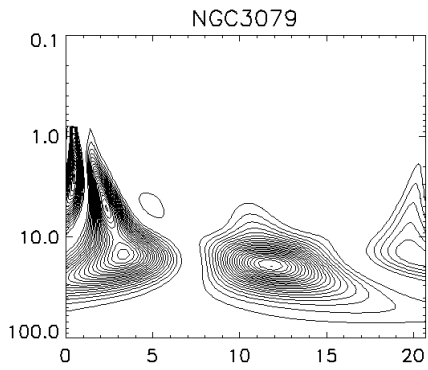
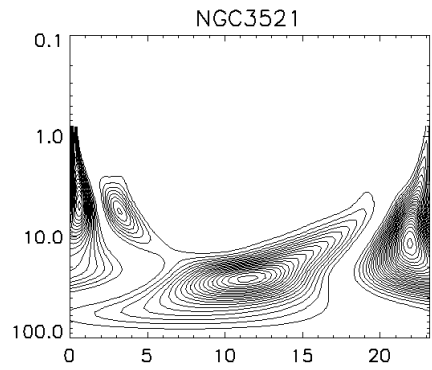
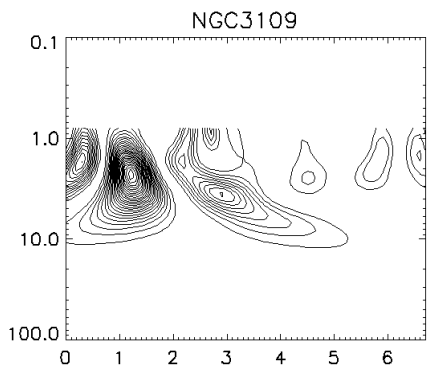
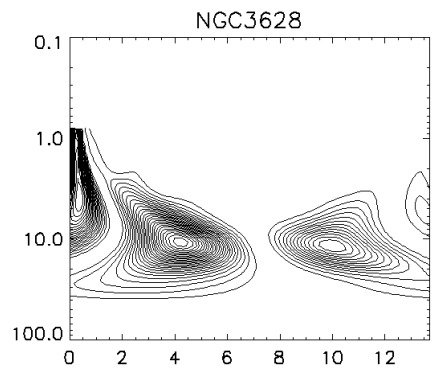

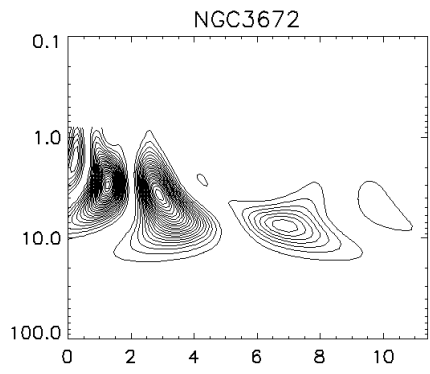
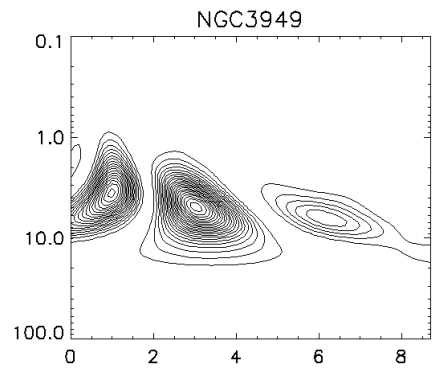
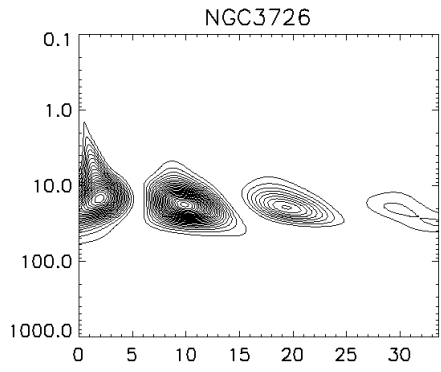
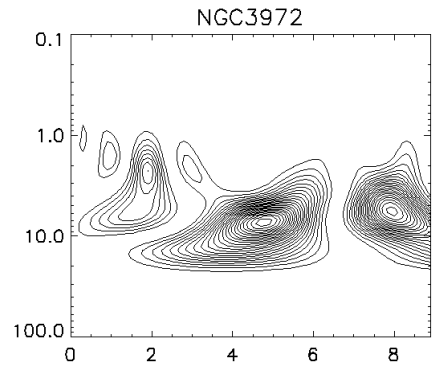
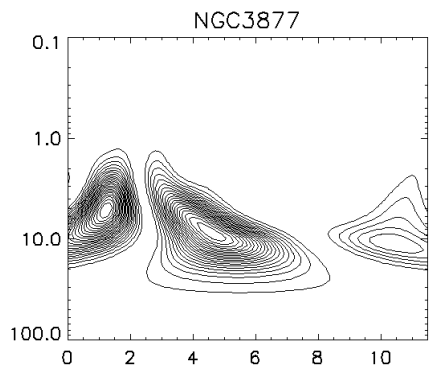
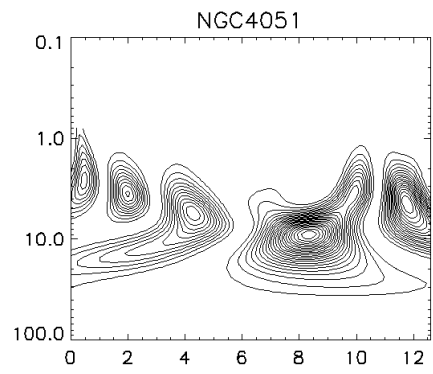
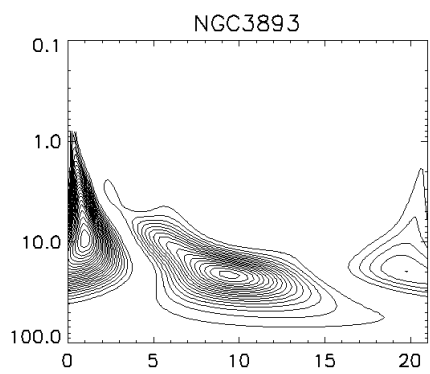
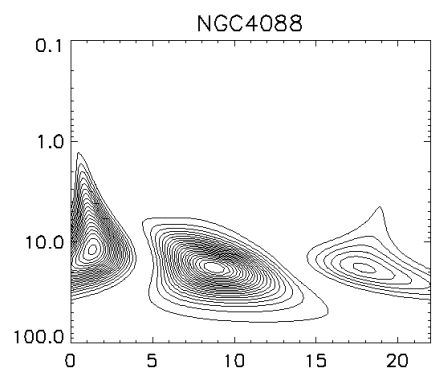

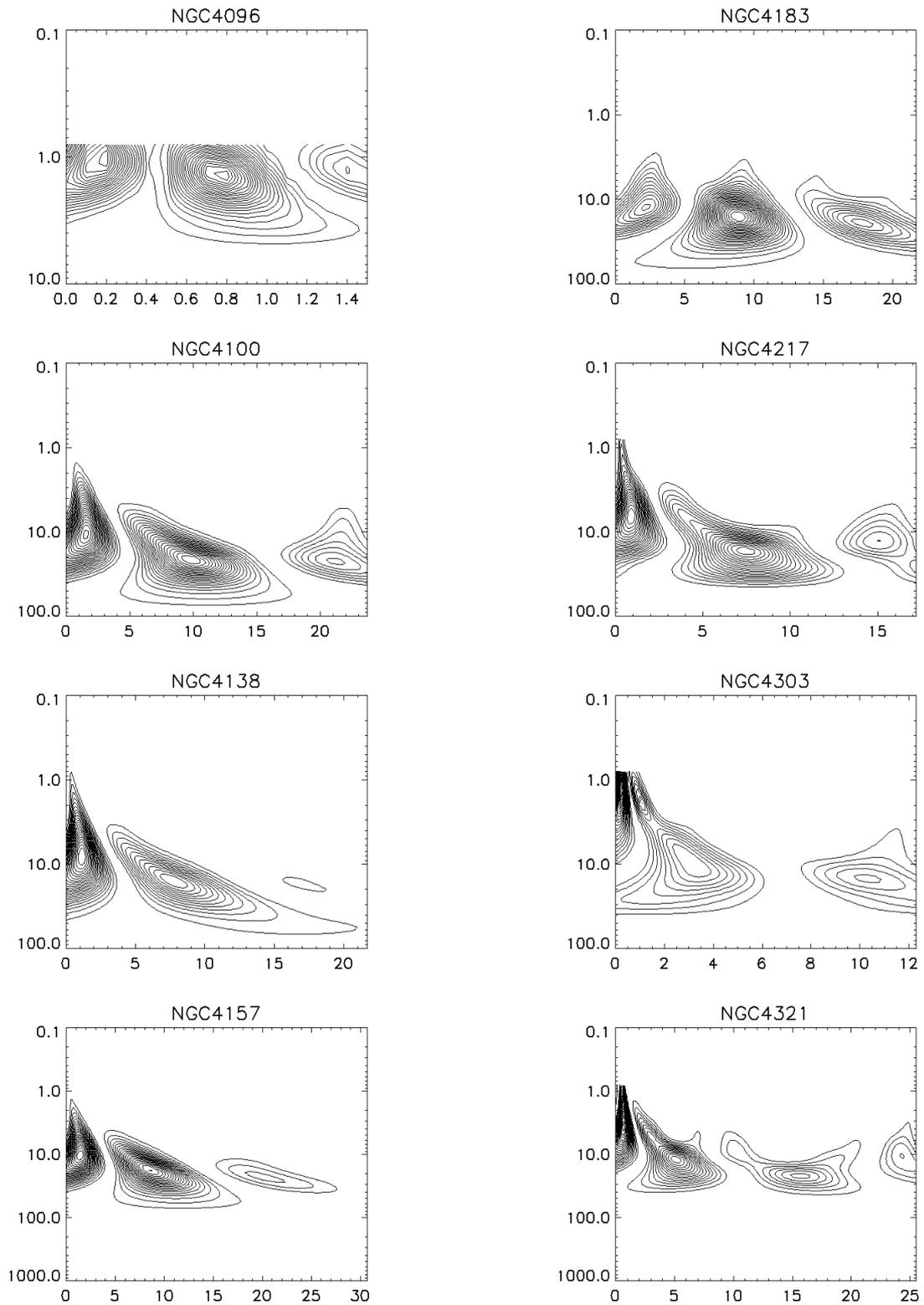

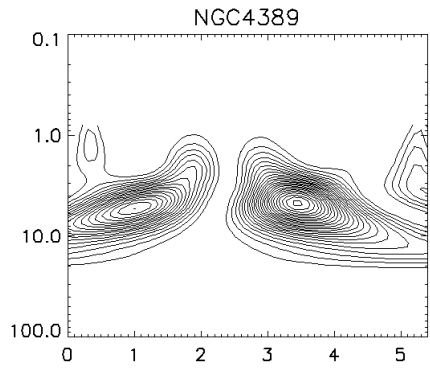
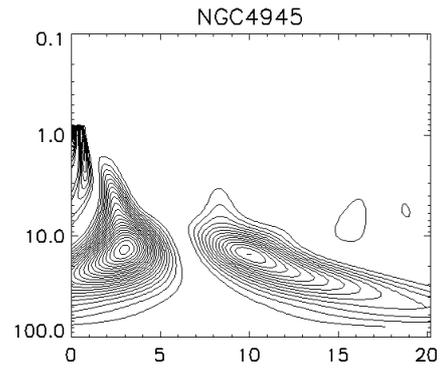
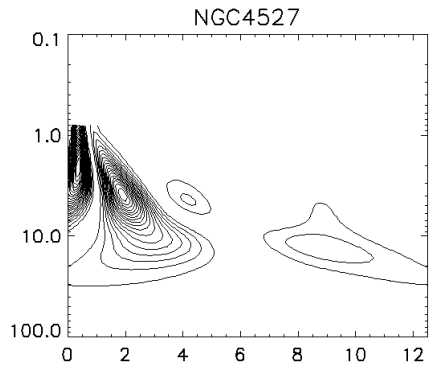
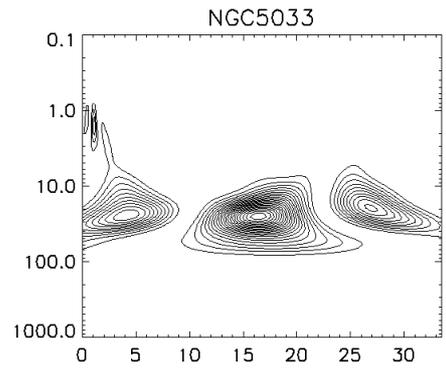
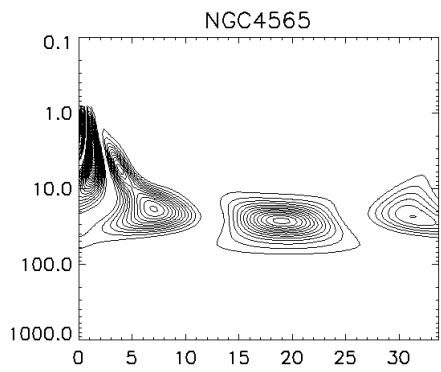
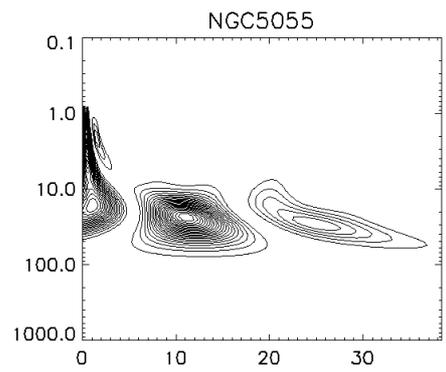
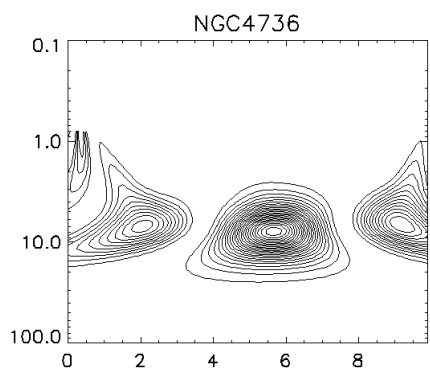
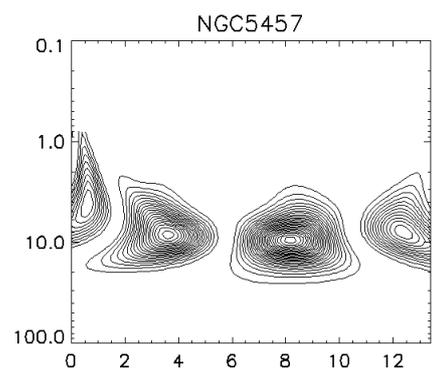

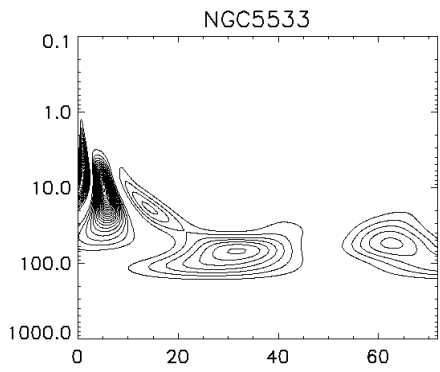
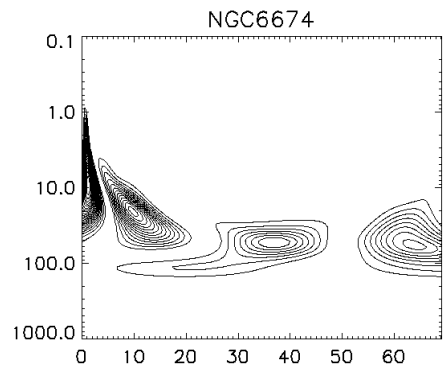
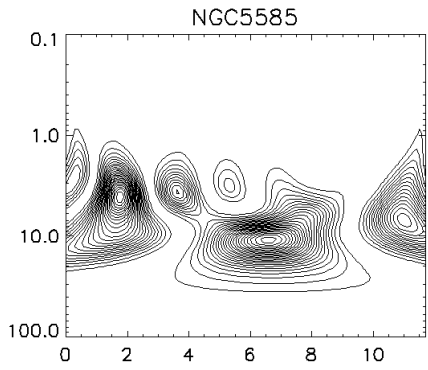
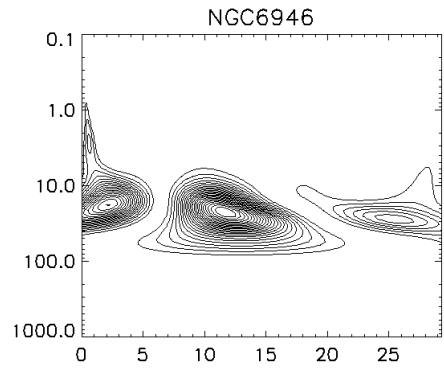
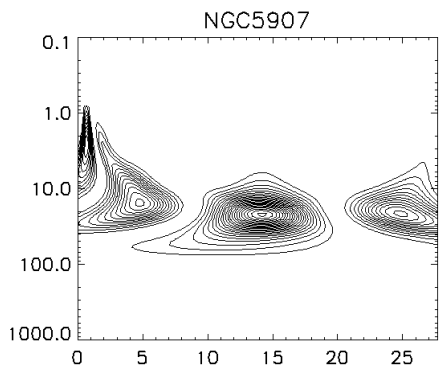
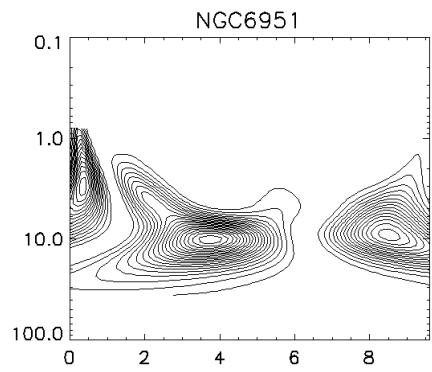
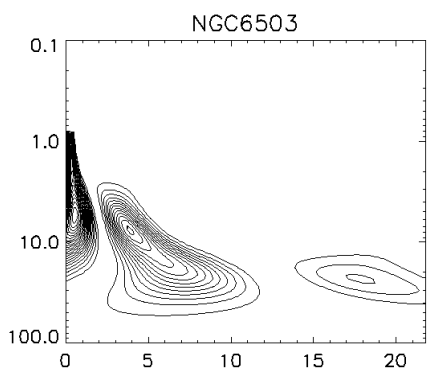
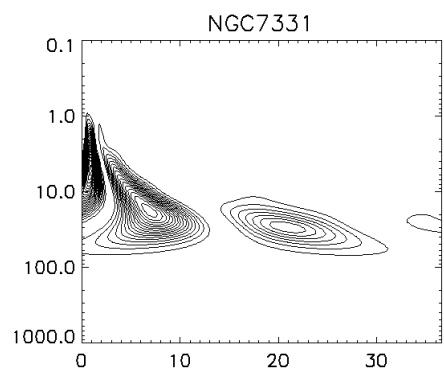

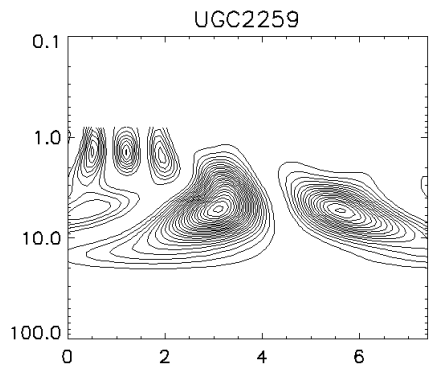
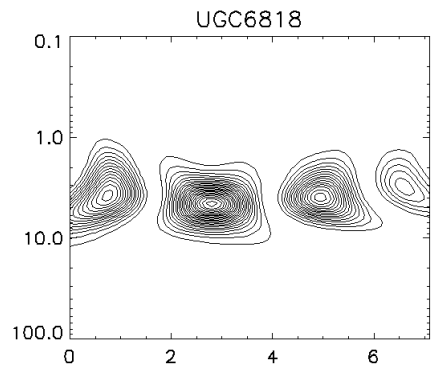
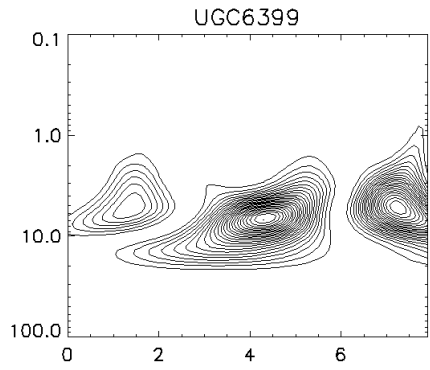
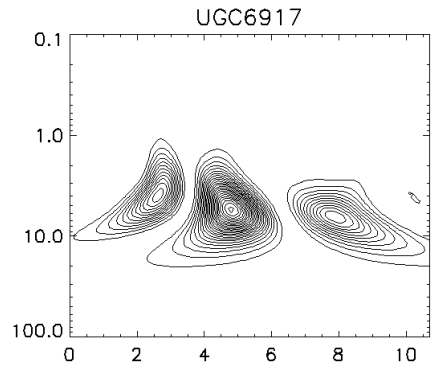
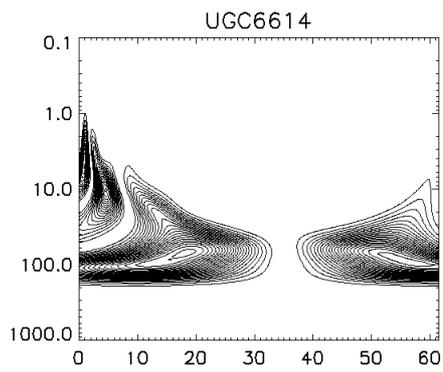
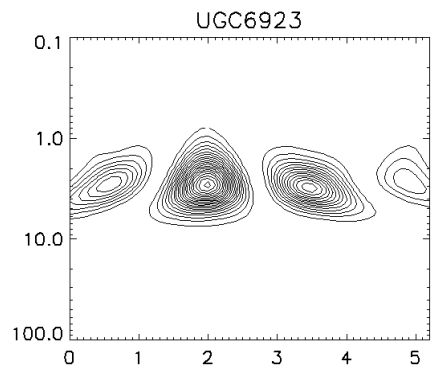
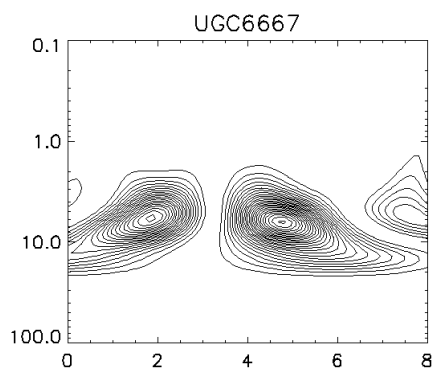
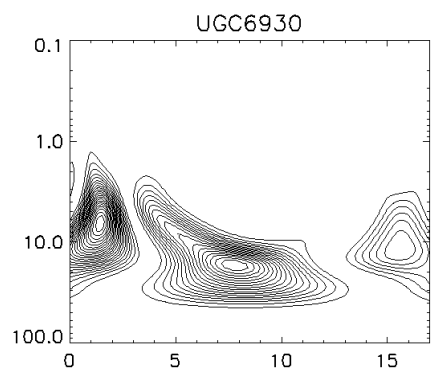

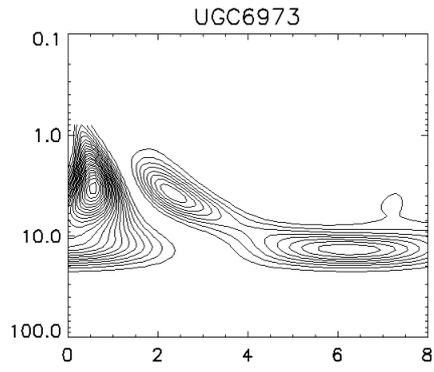
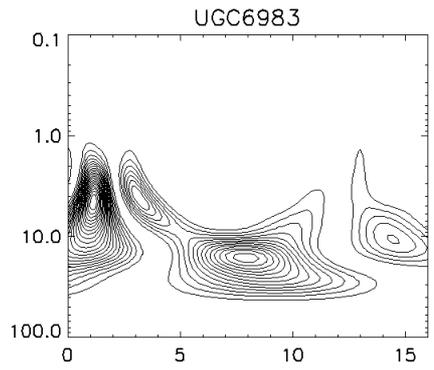
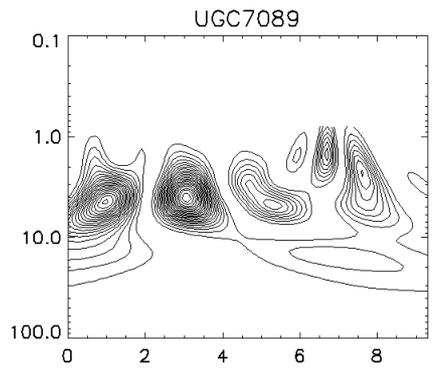